\newif\ifMNRAS
\MNRASfalse % use this line for ApJ

\PassOptionsToPackage{pdfpagelabels=false}{hyperref} 

\ifMNRAS
    \documentclass[fleqn,usenatbib,useAMS]{mnras}
    \usepackage{graphicx}
\else
    \documentclass[twocolumn]{aastex63}
\fi
\usepackage{acronym}
\usepackage{hyperref}
\usepackage{amsmath,amsfonts,amssymb}
\usepackage{mathrsfs}
\usepackage{mathtools}
\usepackage[utf8]{inputenc}

\usepackage[normalem]{ulem}
\usepackage{xspace}

\usepackage{etoolbox}
\makeatletter % Workaround to only color the year for cite commands
  \patchcmd{\NAT@citex}
    {\@citea\NAT@hyper@{%
      \NAT@nmfmt{\NAT@nm}%
      \hyper@natlinkbreak{\NAT@aysep\NAT@spacechar}{\@citeb\@extra@b@citeb}%
      \NAT@date}}
    {\@citea\NAT@nmfmt{\NAT@nm}%
    \NAT@aysep\NAT@spacechar\NAT@hyper@{\NAT@date}}{}{}

  \patchcmd{\NAT@citex}
    {\@citea\NAT@hyper@{%
      \NAT@nmfmt{\NAT@nm}%
      \hyper@natlinkbreak{\NAT@spacechar\NAT@@open\if*#1*\else#1\NAT@spacechar\fi}%
        {\@citeb\@extra@b@citeb}%
      \NAT@date}}
    {\@citea\NAT@nmfmt{\NAT@nm}%
    \NAT@spacechar\NAT@@open\if*#1*\else#1\NAT@spacechar\fi\NAT@hyper@{\NAT@date}}
    {}{}
\makeatother

\shorttitle{ML in CCSNe}
\shortauthors{Tsang, Vartanyan, and Burrows}
\begin{document}
% Don't change these lines
%\label{firstpage}
%\maketitle
%\ifMNRAS
%\else
%%%%%%%%%%%%%%%%%%%%%%%%%%%%%%%%%%%%%%%%%%%%%%%%%%%%%%%%%%%%%%%%%%%%%%%%
\title{Applications of Machine Learning to Predicting Core-collapse Supernova Explosion Outcomes}
%%%%%%%%%%%%%%%%%%%%%%%%%%%%%%%%%%%%%%%%%%%%%%%%%%%%%%%%%%%%%%%%%%%%%%%%
\correspondingauthor{Benny T.-H. Tsang}
\email{benny.tsang@berkeley.edu} 
\author[0000-0002-6543-2993]{Benny T.-H Tsang}
\affiliation{Department of Astronomy and Theoretical Astrophysics Center, University of California, Berkeley, CA 94720, USA}
\affiliation{Kavli Institute for Theoretical Physics, University of California, Santa Barbara, CA 93106, USA}
\author[0000-0003-1938-9282]{David Vartanyan}
\affiliation{Department of Astronomy and Theoretical Astrophysics Center, University of California, Berkeley, CA 94720, USA}
\author[0000-0002-3099-5024]{Adam Burrows}
\affiliation{Department of Astrophysical Sciences, Princeton University, NJ 08544, USA}
%\fi

\begin{abstract}
Most existing criteria derived from progenitor properties of core-collapse supernovae are not very accurate in predicting explosion outcomes. We present a novel look at identifying the explosion outcome of core-collapse supernovae using a machine learning approach. Informed by a sample of 100 2D axisymmetric supernova simulations evolved with F{\sc{ornax}}, we train and evaluate a random forest classifier as an explosion predictor. Furthermore, we examine physics-based feature sets including the compactness parameter, the Ertl condition, and a newly developed set that characterizes the silicon/oxygen interface.
With over 1500 supernovae progenitors from 9$-$27 M$_{\odot}$, we additionally train an auto-encoder to extract physics-agnostic features directly from the progenitor density profiles.
We find that the density profiles alone contain meaningful information regarding their explodability. Both the silicon/oxygen and auto-encoder features predict explosion outcome with $\approx$90\% accuracy. 
In anticipation of much larger multi-dimensional simulation sets, we identify future directions in which machine learning applications will be useful beyond explosion outcome prediction. 

\end{abstract}

\keywords{ \emph{Unified Astronomy Thesaurus concepts}: Core-collapse supernovae (304); Astronomy data analysis (1858); Astrostatistics Techniques (1886); Classification (1907); Random forest (1935); Convolutional neural networks (1938)}

\section{Introduction}
\label{sec:int}

% Key points to cover:
%% ML getting popular in astrophysics
%% Applications are diverse, across different subfield, give some examples.
%% State that applications in the pipeline of CCSN modeling is scarce. 
%% State broad potential applications
%% State the scope of current work: proof-of-concept using the classification task.

Machine learning (ML) has become an integral part of astrophysics research in the recent decade \citep{BB10,Ivezic2014,Fluke20}.
In essence, ML systems are computational tools that are efficient in assimilating complex probability distributions. These distributions are ubiquitous in both observational and theoretical astronomy. For example, characteristic separation of data samples in the image domain has facilitated reliable classification of galaxies \citep{Aniyan17,Cheng20}. Similar success has been achieved in the time domain for variable star classification \citep{Naul18,vanRoestel21}. In addition, identification of outliers from data clusters of known types, a technique known as novelty or anomaly detection, enables the discovery of previously unknown objects and new classes of objects \citep{Williamson19,Villar20,Tsang19,Ishida21,Malanchev21,Bengyat22}. 

Beyond classification and detection, one can regard multi-physics simulation products themselves as the complex distributions to be learned. Emulations of computationally costly simulations can be generated quickly by sampling new data points in the latent spaces that are trained to embody the fully-fledged simulations \citep{Caldeira19,Mustafa19,Vogl20,Horowitz21}.
Moreover, ML systems are powerful tools for parameter inference, connecting observables to physical parameters that are oftentimes degenerate \citep{Ksoll20,Villanueva-Domingo21,Villanueva-Domingo22}. 
Parameters can even be distributions themselves, e.g., the equation of state of neutron stars \citep{Krastev22}, the tensor closure for neutrino transport \citep{Harada22}, and turbulence closures for sub-grid modeling \citep{Karpov22}.

However, entirely lacking is the application of ML techniques to predicting core-collapse supernovae (CCSNe) explosion outcomes. CCSNe simulations are computationally expensive endeavors in both human and machine terms, and thus are a ripe opportunity for ML application. The explosion mechanism of CCSNe through the neutrino-heating mechanism has been studied as a computational problem for more than half a century (\citealt{1966ApJ...143..626C, 1985ApJ...295...14B}), through both detailed computational simulations and much cruder prescriptive methods (e.g., imposing a thermal bomb, driving a piston, or other rudimentary prescriptions). Only in the last decade have multi-dimensional simulations become the mainstay, with various groups performing scores of two-dimensional axisymmetric computations and tens of three-dimensional simulations. Though population suites of CCSNe have been evolved in 2D (\citealt{burrows2018, radice2017b,vartanyan2018a, summa2016, Ertl16,2022ApJ...924...38K, 2021Natur.589...29B}) and, more selectively, 3D (\citealt{vartanyan2018b,burrows_2019,burrows2020, nagakura2019b,  glas2019,oconnor_couch2018b, summa2018, 2020ApJ...896..102K, 2021MNRAS.503.4942O,2022MNRAS.510.4689V}), developing hundreds, let alone thousands, of 3D simulations may not be feasible in the coming decade.

To circumvent this limitation, and in order to explore the explosion landscape by progenitor for final explosion energies, observational signatures, and nucleosynthetic compositions, various groups have developed CCSNe population studies using simplified prescriptions in reduced dimensions. Different such approaches include analytical approximations of proto-neutron star cooling (\citealt{2012ApJ...757...69U}), PUSH (\citealt{PUSH1,2021ApJ...921..143C}), simple pistons (\citealt{swbj16}), and spherically symmetric turbulence models (STIR, \citealt{2020ApJ...890..127C, 2019ApJ...887...43M}), often calibrated to SN1987a and the Crab and comparing the derived explosion outcomes with various formulated predictions (e.g., the antesonic condition, \citealt{2012ApJ...746..106P,2018MNRAS.481.3293R}, the Ertl criterion, \citealt{Ertl16}; a semi-analytical pre-SN parametrization, \citealt{2016MNRAS.460..742M}).

These methods rely on simplifying approximations for both explosion modeling and explosion prediction. In light of this, the motivation of our paper is to present a summary overview of potential ML approaches to CCSNe outcome prediction as a proof-of-concept of the eventual goal $-$ developing ML techniques, trained on the results of extant multi-dimensional simulations, to predict explosion outcome while circumventing costly detailed simulations. Our intent here is not to be comprehensive, but rather to present a sample of the applicable methods and to galvanize the use of these techniques more broadly in the community. 
We wish to highlight the versatility and potential future use of ML, and identify potential difficulties and obstacles.

In Section \ref{sec:methods}, we describe our methodology including the simulated dataset (Section \ref{sec:datasets}), the various physics-based explosion conditions (Section \ref{sec:physics_features}), an unsupervised feature extraction approach used to derive physics-agnostic explosion criteria (Section \ref{sec:unsupervised_features}), a baseline random forest (RF) classifier used as an explosion outcome predictor (Section \ref{sec:classifier}), and a semi-supervised label propagation approach (Section \ref{sec:lp}). In Section \ref{sec:results}, we present the results comparing the accuracy of the various features in predicting explosion outcomes. We summarize our conclusions in Section \ref{sec:conc} and identify future directions in Section \ref{sec:future}.

\section{Methods}
\label{sec:methods}

Our goal is to survey various machine learning approaches in tandem with a selection of explosion criteria to study their value in predicting explosion outcome ab-initio. These explosion outcome predictors are trained and tested on a suite of 100 2D axisymmetric CCSNe simulations run with the radiation-hydrodynamic code F{\sc{ornax}}.
F{\sc{ornax}} \citep{skinner2019} is a multi-dimensional, multi-group code constructed to study CCSNe. It features an M1 solver \citep{2011JQSRT.112.1323V} for neutrino transport with detailed neutrino microphysics and an approximation to general relativity \citep{marek2006}.

\begin{figure*}
\centering
{
\includegraphics[width=\columnwidth]{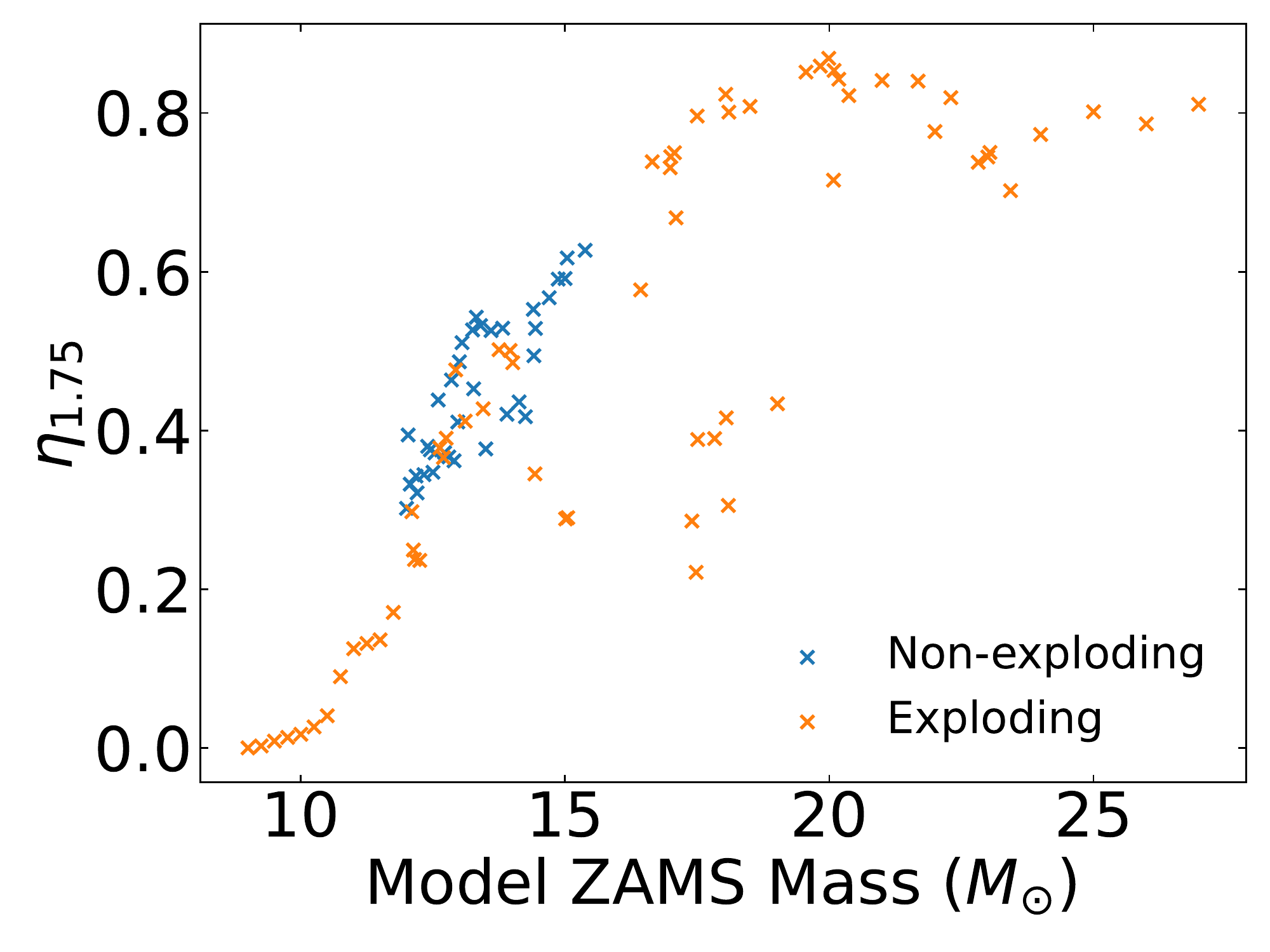}
\includegraphics[width=\columnwidth]{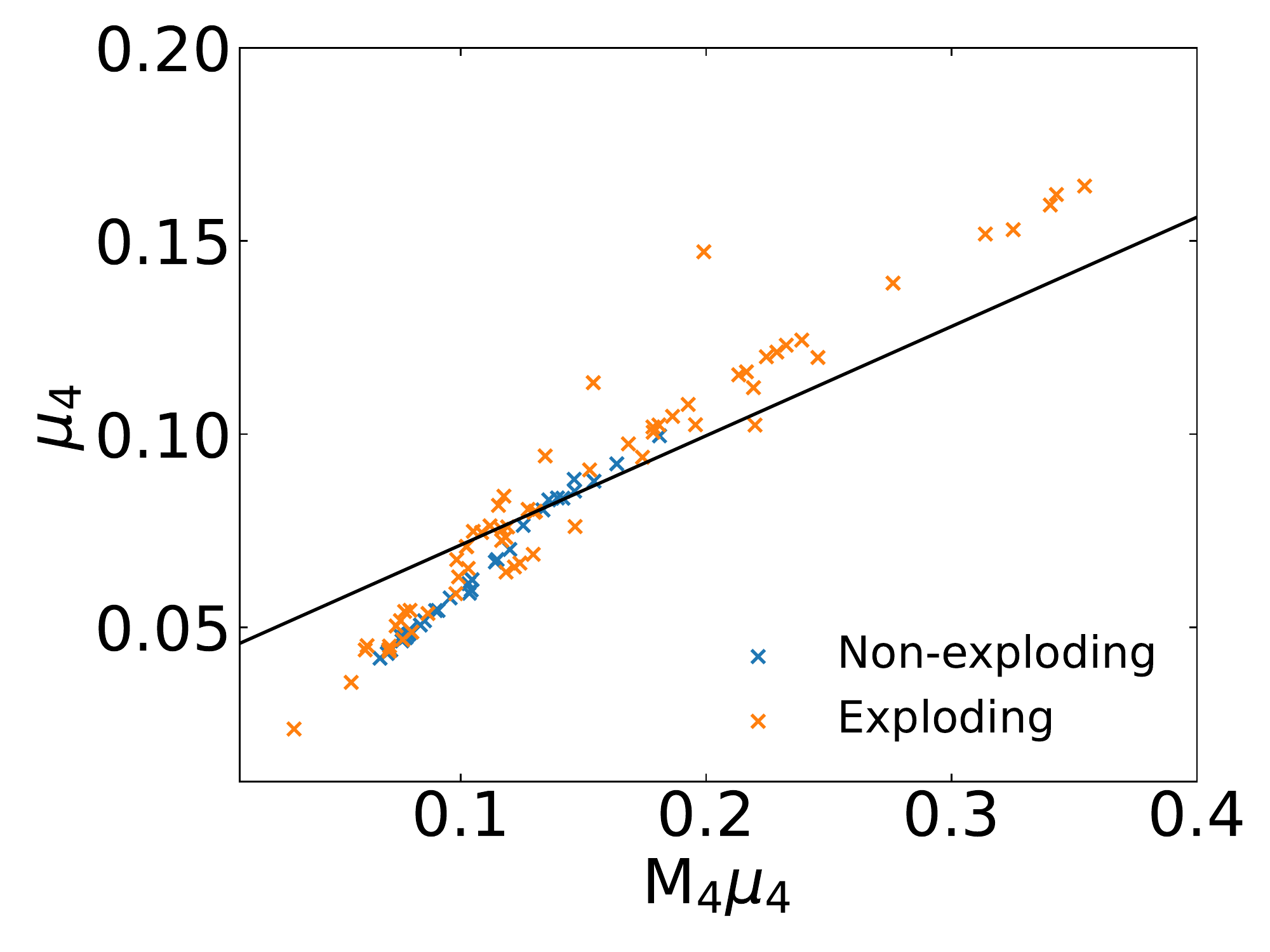}
}
\caption{The compactness $\eta_{1.75}$ vs ZAMS mass (\textbf{left}) and the Ertl parametrization (\textbf{right}) for the 100 labeled models evolved in 2D. Blue crosses indicate non-exploding models, and orange crosses exploding models. The putative separation curve in the Ertl parametrization does not separate explosion outcomes reliably, with exploding models both above and below the line, although non-exploding models are almost entirely below the line (\protect{\citealt{2022arXiv220702231W}}). Note the clustering of non-exploding models in both figures, particularly between 12 and 15 M$_{\odot}$ with different outcomes vis-\'a-vis explosion for a given physical parameter (see also \protect\citealt{sukhbold2018}), perhaps indicative of a third dimension necessary to break the degeneracy. }
\label{fig:Ertl_eta_plane}
\end{figure*}

\subsection{Datasets}
\label{sec:datasets}
We selected a subset of 100 initial progenitor models from 9$-$26.99 $M_{\odot}$ to evolve in 2D-axisymmetry (Vartanyan et al., in prep.) using F{\sc{ornax}} for typically one second after core bounce to ascertain their explodability (discussed in more detail in \citealt{2022arXiv220702231W}). These models were evolved with neutrino heating as the explosion mechanism, absent rotation and magnetic fields. The models had a resolution of 1024$\times$256 in $r$, $\theta$ with outer radii extending from 30,000 km for the lower mass stellar progenitors and to 100,000 km for the most massive progenitors. These models were chosen to be representative as much as possible of the Salpeter initial mass function. They were selected to span broadly the distribution in density profiles, compositional interfaces, compactness and $\mu_4$/$M_4$ (discussed below).
We categorize explosion as a run-away shock radius within the simulation time. Of the 100 models, 64 explode and 36 did not. 
These 100 models with known explosion outcomes based on the 2D simulations will be referred to as the \emph{labeled} dataset.
Our 100 models were selected from the newest stellar progenitor models in \citealt{swbj16,sukhbold2018}. 
The compilation contains 12 progenitors in the mass range of 9$-$11.75\,$M_{\odot}$ in increments of 0.25\,$M_{\odot}$ (from \citealt{swbj16}), and 1500 progenitors in the range of 12$-$26.99\,$M_{\odot}$ in increments of 0.01\,$M_{\odot}$, for a grand total of 1512 stellar progenitors.
The 1412 progenitor models that were not evolved in F{\sc{ornax}}, and therefore do not have known explosion outcomes, are referred to as the \emph{unlabeled} dataset.
All the models studied were evolved as single-star progenitors, absent binary effects. We note that we are limited by the dataset size of this ML exercise, as well as by the complexity of the physical phase-space explored.

\subsection{Physics-based Features}
\label{sec:physics_features}

Due to the limited number and high dimensionality of the progenitor models, explosion outcome predictors in the form of binary classifiers cannot be well-trained using the raw stellar profiles as inputs. Instead, parameters of much lower dimension, known as \emph{features}, are obtained to represent the distinctive characteristics of the models in a process known as \emph{feature extraction}.
Multiple attempts have been made to identify such explosion conditions, often ab-initio, that can serve to predict CCSNe outcome (e.g., \citealt{2011ApJ...730...70O, 2012ApJ...746..106P, dolence_2015, Ertl16, 2016MNRAS.460..742M, summa2018, 2022MNRAS.515.1610G}). These derive in heritage from some variation on the concept of a critical condition \citep{burrowsgoshy1993}, which suggests a relation between neutrino luminosity and mass accretion at the shock, above which unabated shock expansion concludes in explosion. Below, we summarize three types of explosion metrics whose utility in predicting explosion outcomes we explore with our ML approaches. We focus on compactness and the Ertl condition because of their widespread use, their relative simplicity, and their ab-initio nature. We also target an additional feature $-$ the role of the silicon-oxygen compositional interface.

\subsubsection{Compactness}

The compactness parameter characterizes the core structure and is defined as \citep{2011ApJ...730...70O}:

\begin{equation}
\xi_M= \frac{M/M_{\odot}}{R(M)/1000\, \mathrm{km}}\,,
\end{equation}
where the subscript $M$ denotes the interior mass coordinate at which the compactness parameter is evaluated. For our purposes, we evaluate the compactness parameter $\xi_{1.75}$ at $M$ = 1.75 M$_{\odot}$, generally encompassing the Si/O interface for many the progenitor models. The compactness is often used as an ab-initio explosion condition because it depends only on the progenitor properties. While higher compactness is correlated with higher luminosities, accretion rates, and remnant masses \citep{oconnor2013},  compactness does not readily lend itself as an explosion condition, and suggestions that explosion is inhibited above a certain compactness parameter are false \citep{burrows2020} (with the exception that massive models may initially drive a successful shock, but then later implode into a black hole due to the large gravitational binding energy). 
We plot in the left panel of Figure \ref{fig:Ertl_eta_plane} the distribution of compactness versus the zero-age main sequence (ZAMS) mass of the labeled dataset, with exploding models indicated in orange and non-exploding in blue. For most progenitors the mass at which $\xi_M$ is calculated usually encompasses the Si/O interface entropy and density jump, and this is discussed below.  

\subsubsection{Ertl Parameter}

The Ertl condition for explosion \citep{Ertl16} is another ab-initio explosion condition. It identifies a $\mu_4$ and $\mu_4 \times M_4$ space, where $\mu_4$ is a measure of the slope of the mass density at an entropy of four (per baryon per Boltzmann's constant) and $M_4$ is the interior mass at that entropy. This approximately corresponds to the location of an entropy/density jump at the Si/O interface. The Ertl condition purports to be a statement of criticality, with $\mu_4$ and $\mu_4 \times M_4$ relating indirectly to $L_{\nu}$ and $\dot{M}$, the neutrino luminosity and the mass accretion rate. We show in the right panel of Figure \ref{fig:Ertl_eta_plane} the Ertl curve suggested to separate explosion and non-explosion, overplotted with the results of our 100 2D simulations (see also \citealt{2022arXiv220702231W}). We note the poor agreement between our simulation results and the Ertl prediction, and comment on this more in Section \ref{sec:results}.

\begin{figure}
\centering
{
 \includegraphics[width=\columnwidth]{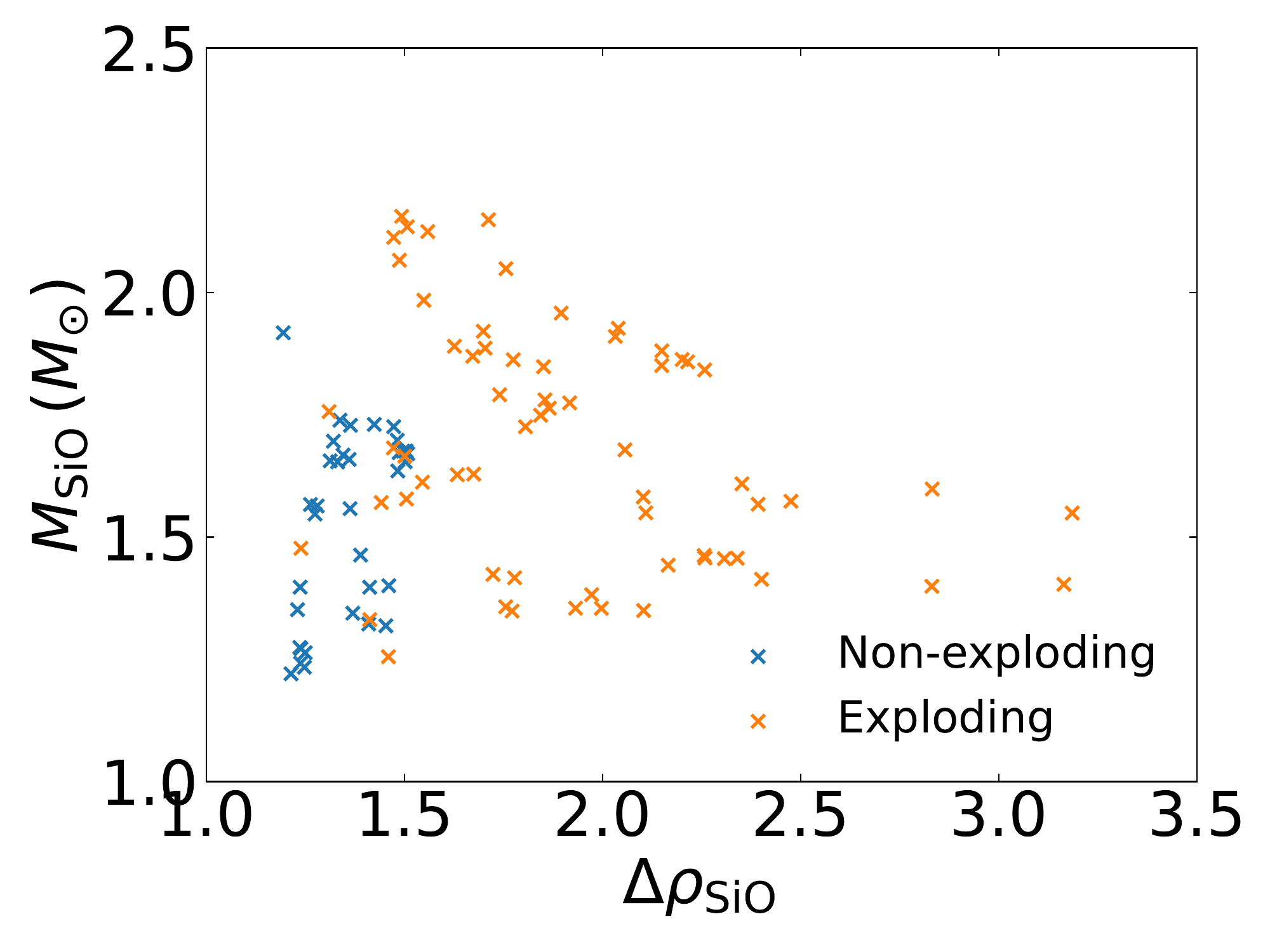}
}
\caption{The Si/O interface parameter distribution of the 100 labeled models evolved in 2D. Blue crosses indicate exploding models, and orange crosses non-exploding models. Note the clumping in phase space for weaker interfaces located deeper in.}
\label{fig:SiO_param_plane}
\end{figure}

\subsubsection{Si/O Interface Parameters}

Lastly, we posit a physically-motivated explosion condition that looks at prominent density interfaces (often the Si/O interface, \citealt{2022arXiv220702231W}) whose accretion by the shock surface can revive a stalled shock into successful explosion \citep{fryer1999,swbj16,burrows2018,vartanyan2018a, ott2018_rel,burrows_2019,2021Natur.589...29B,2021ApJ...916L...5V,Boccioli22}. A sharp drop in density translates into an immediate drop in ram pressure at the shock surface upon encountering this interface, whereas the accretion-powered luminosity interior to the shock is sustained for an advective timescale (\citealt{2022arXiv220702231W}). This drop in ram pressure, while maintaining a higher luminosity, promotes explosion and may be key to explosion for massive stars. Lower-mass models of $\approx$9$-$10 $M_{\odot}$ may explode simply on the virtue of their very steep density profiles.  We identify the location in mass coordinate $M_{\rm SiO}$ and the magnitude of the density jump across such interfaces $\Delta\rho_{\rm SiO}$ in all 1512 models in the full progenitor dataset. For each of the models studied here, we identified the Si/O or other prominent interface by looking for the steepest drop in density in the stellar progenitor profile exterior to the iron core. The stellar density may drop by as much as 2$-$3 times over less than 0.01 M$_{\odot}$.  

The extraction of the interface features can be complicated by the presence of multiple, fragmented burning shells (see also \citealt{2021ApJ...916L...5V, laplace2021} for a similar conclusion regarding the density profiles of binary stars). \citet{sukhbold2018} identify multiple burning shells during late-stage stellar evolution, where the physics is poorly resolved and the results prone to stochasticity (see also \citealt{2022arXiv220702231W}). Merging of the two shells into a single steeper shell will produce a more prominent density drop  conducive to successful core-collapse explosions. We plot in Figure \ref{fig:SiO_param_plane} the Si/O interface mass coordinate versus the magnitude of the density drop from the labeled dataset.
We see a nonuniform distribution of both compactness (in Figure \ref{fig:Ertl_eta_plane}) and the Si/O interface with clustering and multiple branches (see also \citealt{sukhbold2018,2016MNRAS.460..742M,2022arXiv220702231W}), as well as multi-valued outcomes (explosion and not) in a small range of the plotted phase space. Using the Si/O interface yields a clearer delineation between explosion and non-explosion than compactness, but in both cases we see degeneracy in the outcome for the given putative explosion criteria.

\begin{figure*}
\centering
{
 \includegraphics[width=\textwidth]{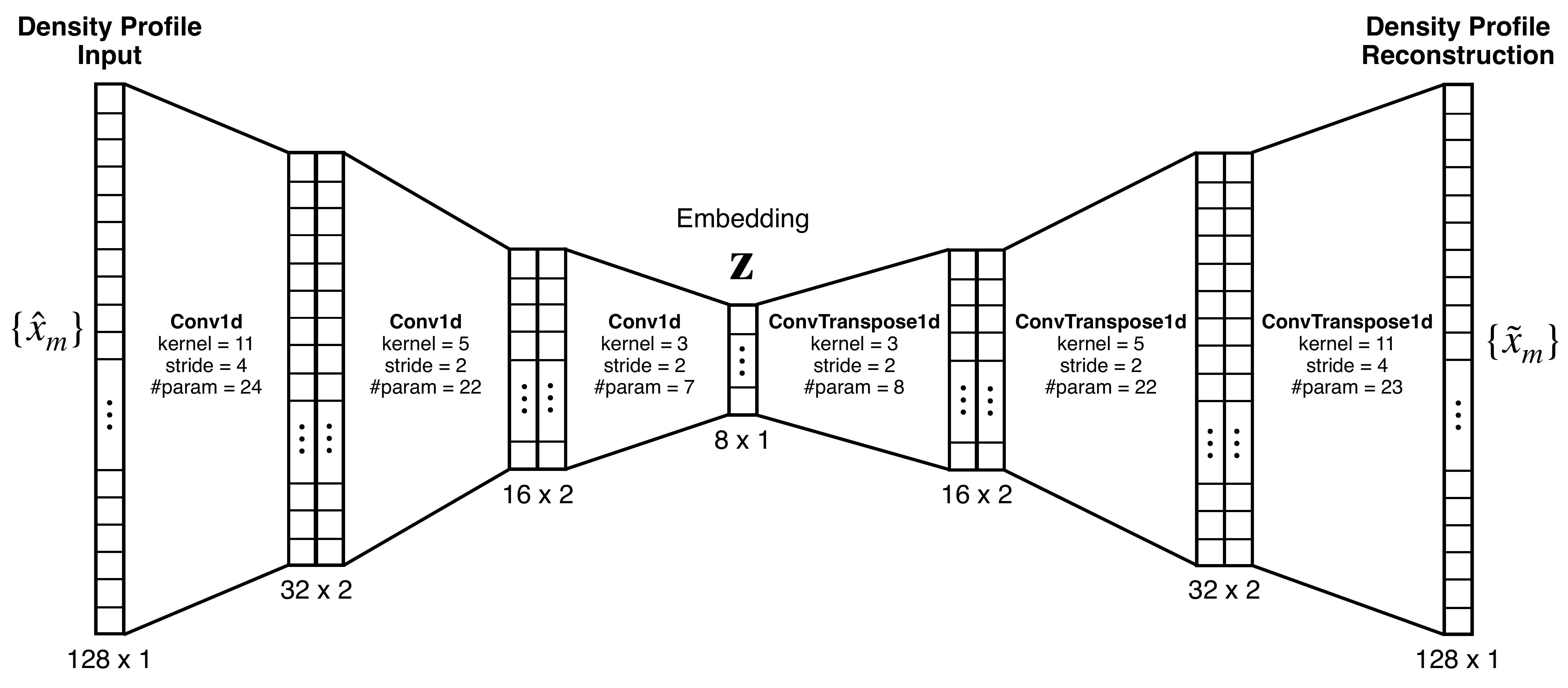}
}
\caption{Schematic diagram of the auto-encoder's neural network architecture. It employs three basic \textsc{Conv1d} layers as the encoder, converting the input density profile sequence \{$\hat{x}_{m}$\} into the reduced-dimension embedding vector $\mathbf{z}$. The decoder similarly utilizes three \textsc{ConvTranspose1d} layers, trained to produce a reconstruction of the input density profile \{$\tilde{x}_{m}$\}. The hyperparameters, number of trainable parameters, and the dimension for each layer are annotated.} 
\label{fig:AE_schematic}
\end{figure*}

\subsection{Unsupervised Feature Extraction}
\label{sec:unsupervised_features}

The physics-based feature sets described in Section \ref{sec:physics_features} are not always very effective. Modern auto-encoder neural network architectures offer an alternate data-driven, physics-agnostic approach to extracting relevant features in an unsupervised manner. Auto-encoders consist of two main components: an encoder and a decoder. The encoder is designed to take an input vector and convert it into a feature vector of much lower dimension. The decoder, on the other hand, attempts to reconstruct the input from the feature vector. By training the encoder-decoder pair to match the input and the reconstruction, the auto-encoder learns to capture the important information in the input without human intervention. Example usage in astronomy includes variable star \citep{Naul18,Tsang19} and galaxy \citep{Portillo20} classification, anomaly detection for supernova light curves \citep{Villar20}, detection of strong lensing features in images \citep{Cheng20}, and the denoising of radio images \citep{Gheller22}. Auto-encoders essentially serve as an apparatus for data compression, allowing ML systems to operate more effectively on the much lighter-weight feature vectors rather than the raw inputs.

Here, we explore the application of auto-encoders to extracting representative features directly from the density profiles of the stellar progenitor models. 
To this end, we implement a basic auto-encoder network in \textsc{PyTorch} \citep{pytorch19}. We adopt three one-dimensional convolutional layers (\textsc{Conv1d}) as the main components of the encoder. The convolutional layers are designed to preserve the spatial information of the mass distribution in the stellar density profiles.
The decoder is constructed using three corresponding \textsc{ConvTranspose1d} layers. The hyperbolic tangent function ($\tanh$) is used as the nonlinear activation function after each \textsc{conv1d} and \textsc{ConvTranspose1D} layer.
After passing through the encoder's final layer, the resultant vector is commonly known as the \emph{embedding}, which is of much smaller dimension than the input sequence. The embedding vectors can be regarded as the reduced-dimension feature vectors that can be used for other downstream tasks. By construction, the $\tanh$ activation function produces embedding vectors $\mathbf{z} \in [-1, 1]^{d_{\rm z}}$, where $d_{\rm z}$ is the embedding dimension. In our case study, we focus on the explosion outcome prediction task, which is set up as binary classification. 
The auto-encoder architecture is presented in the schematic diagram in Figure \ref{fig:AE_schematic}.

To explore the learning capacity of the auto-encoder, we vary the dimension of the embedding vector by adjusting the strides of the convolutional layers, covering embedding dimensions of $d_{\rm z} = 2$ to 32 in factors of 2. The number of total trainable parameters (weights and biases) depends solely on the kernel and filter sizes. Since we do not vary those network parameters, the auto-encoders we use in this work contain a constant number of trainable parameters of 106. The kernel size, stride, and the breakdown of the number of parameters in each layer can be found annotated in Figure \ref{fig:AE_schematic}.
We keep the network size to a minimum to highlight the utility of a simple auto-encoder. 

\begin{figure*}
\centering
{\includegraphics[width=0.49\textwidth]{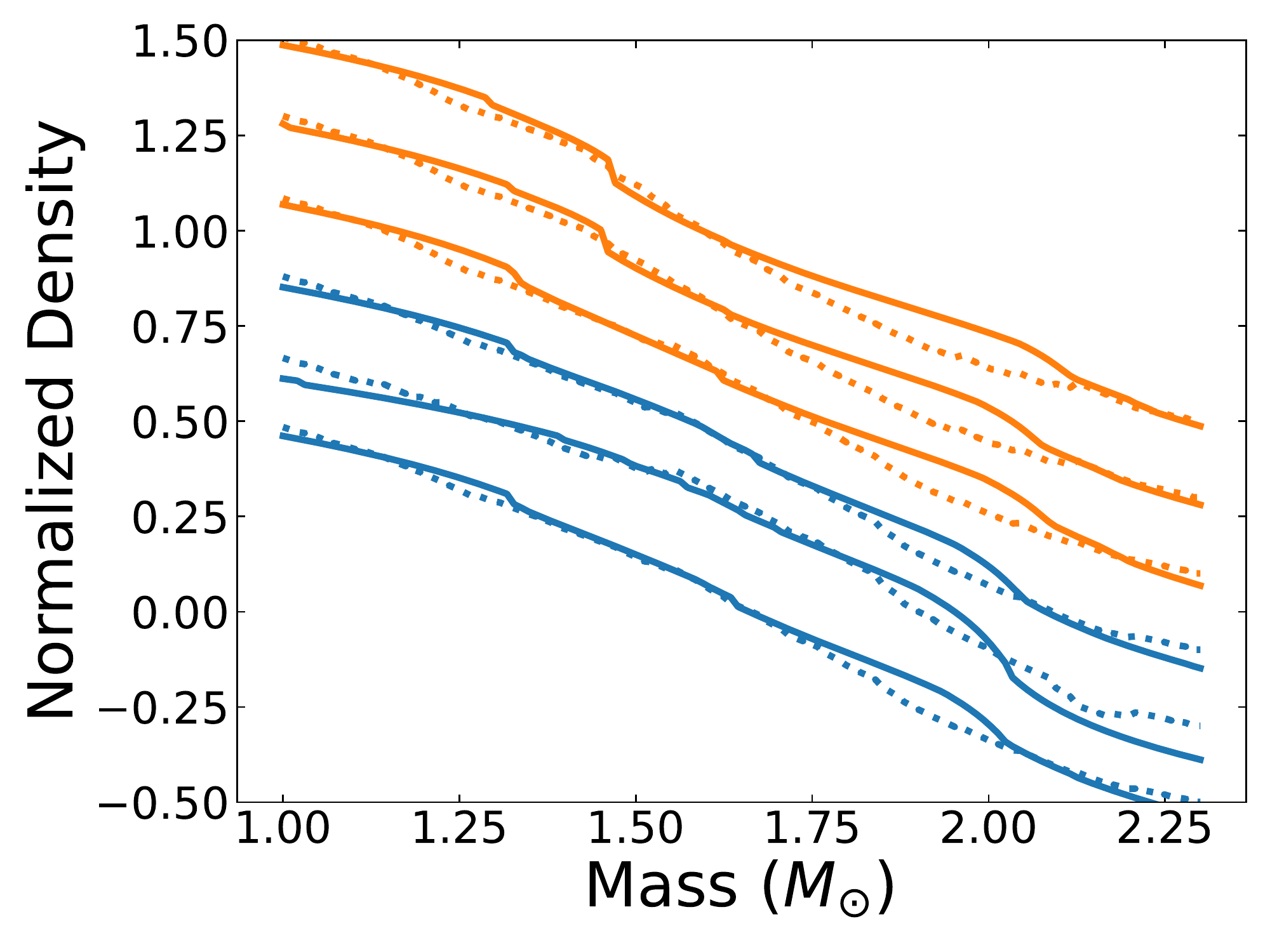}
\includegraphics[width=0.49\textwidth]{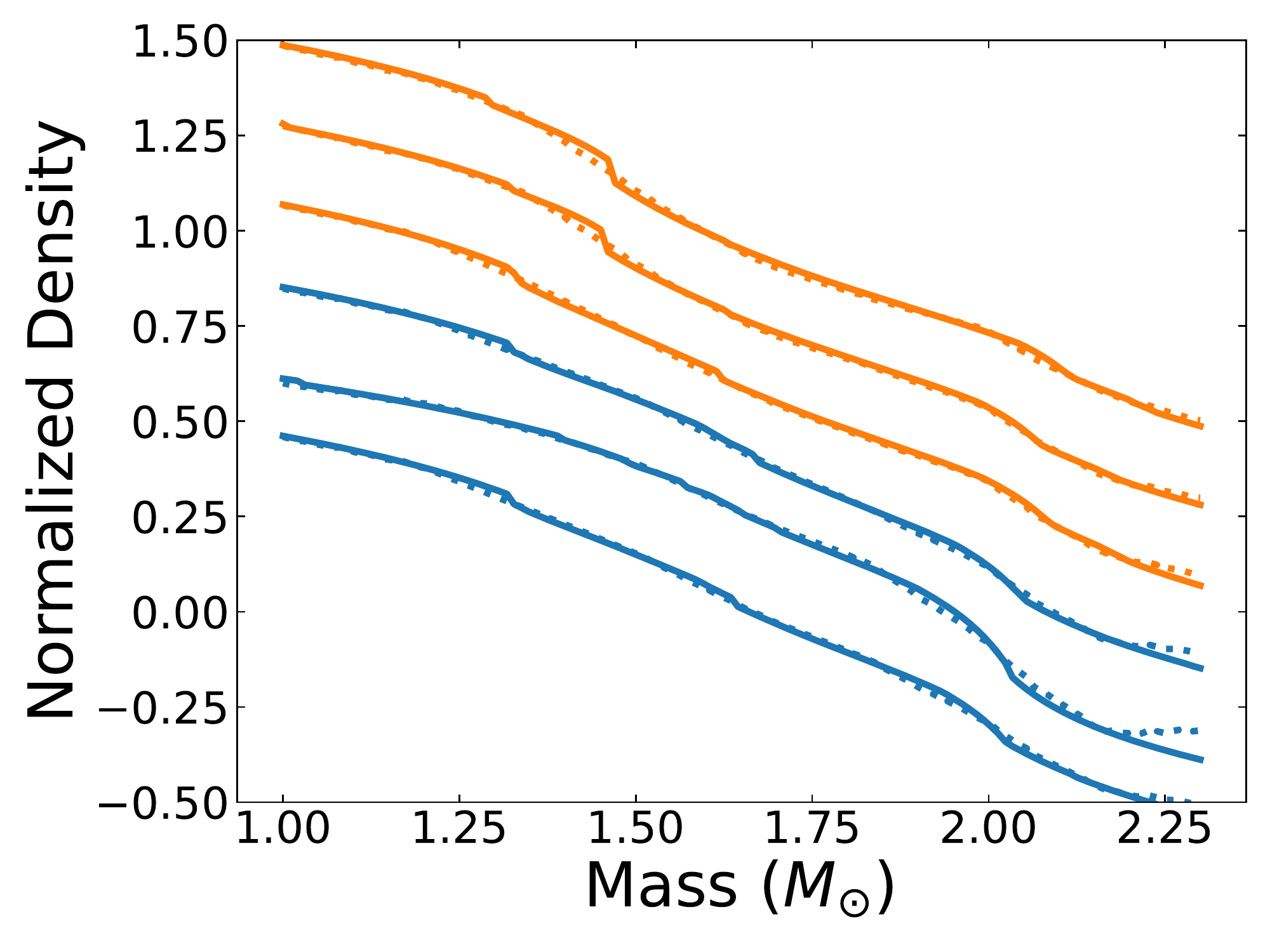}
\includegraphics[width=0.49\textwidth]{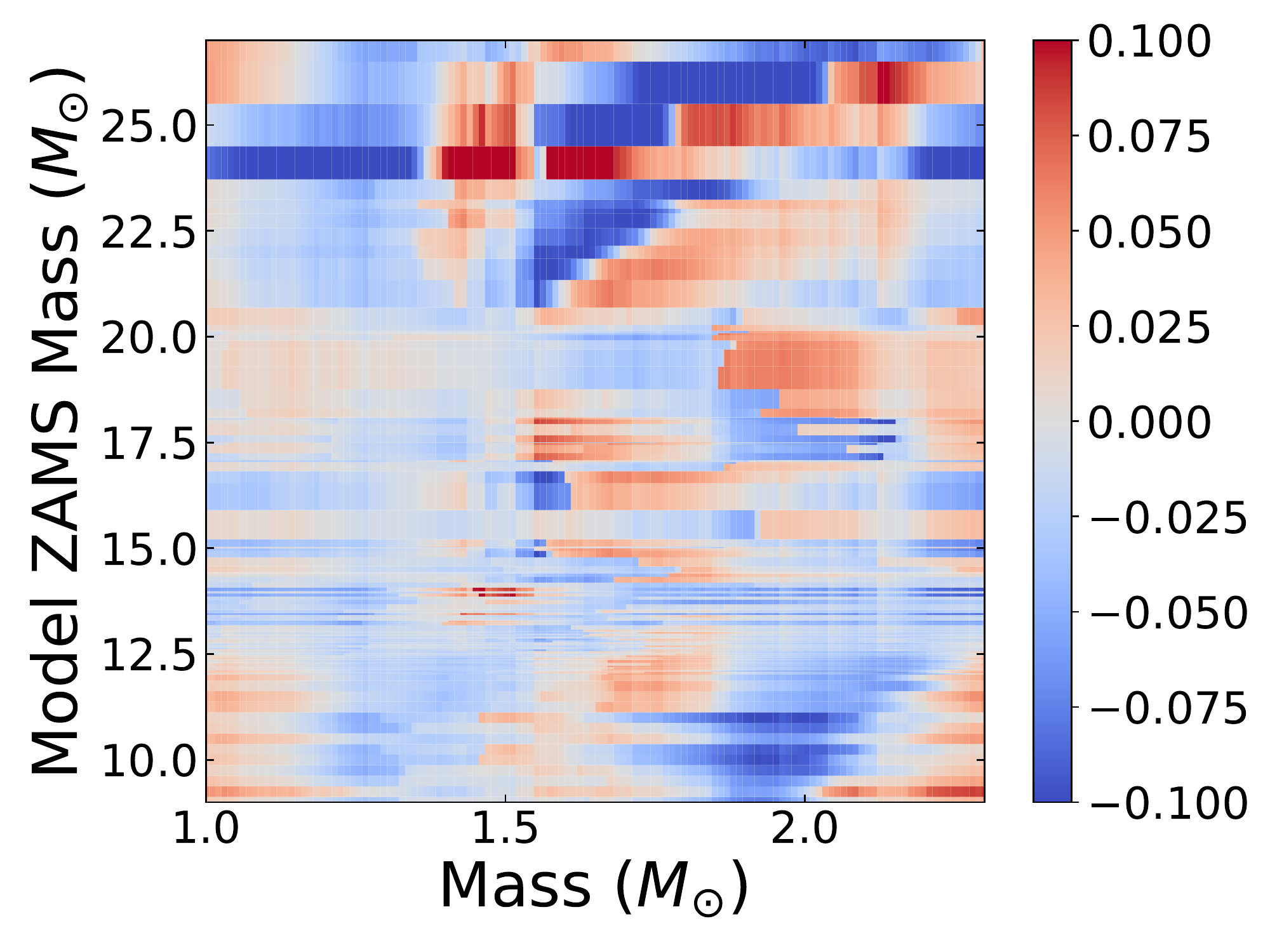}
\includegraphics[width=0.49\textwidth]{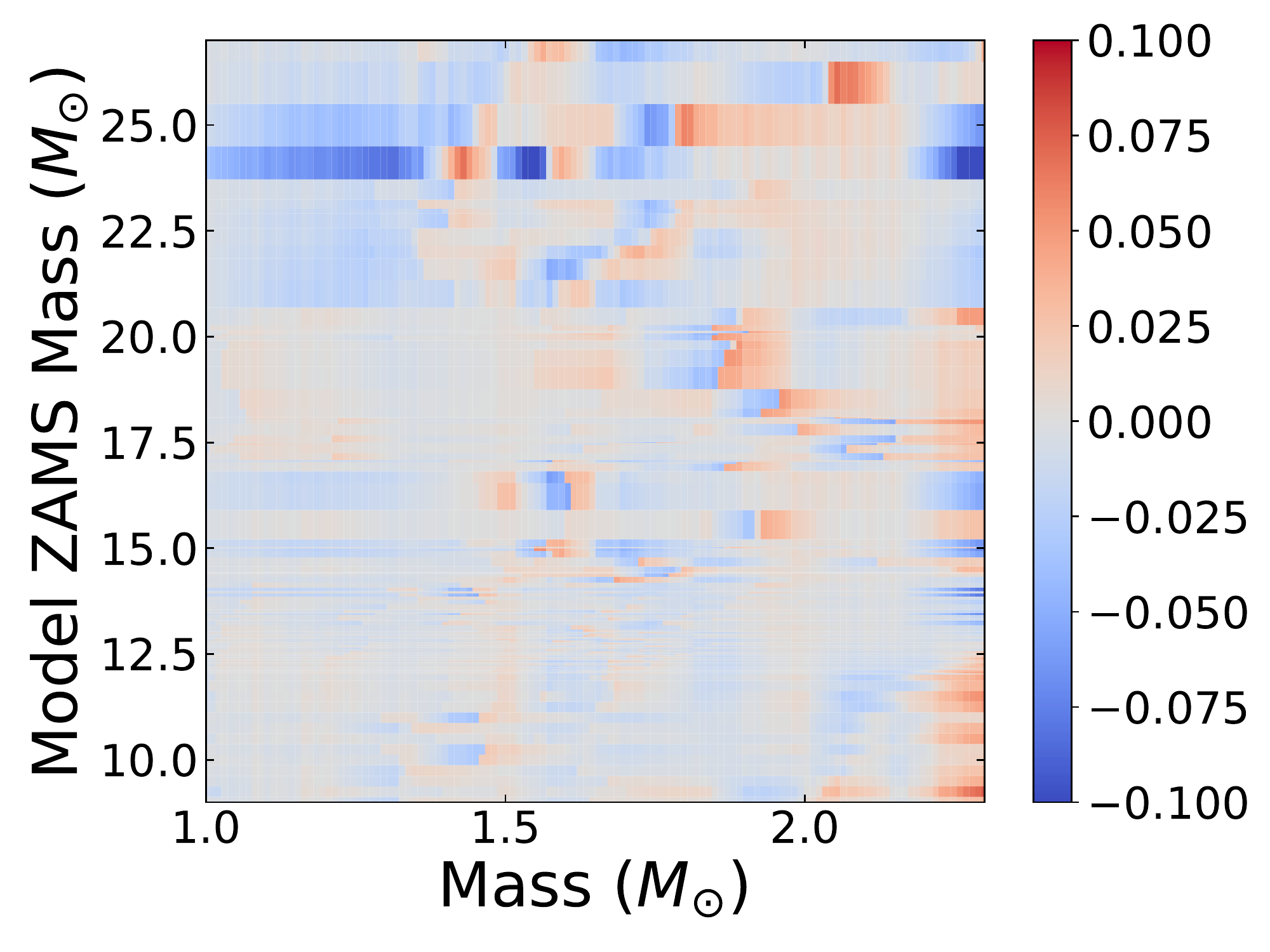}
}
\caption{\textbf{Top}: Examples of the input (solid) and the reconstructed (dotted) density profiles to and from the auto-encoders with embedding dimensions of 2 (left) and 8 (right). Blue (orange) curves correspond to exploding (non-exploding) models. The profiles are plotted with consecutive vertical offsets of 0.1 for better visualization.
\textbf{Bottom}: The reconstruction error (reconstruction $-$ input) of the normalized density profile by the auto-encoders with embedding dimensions of 2 (left) and 8 (right). The vertical axis denotes the ZAMS mass of all 100 models in the labeled dataset. The fiducial model can encode density profiles with errors of $\lesssim$0.05.
}
\label{fig:ae_reconstructions}
\end{figure*}

We use the 1412 unlabeled models as the training set of the auto-encoder models. In other words, the auto-encoder is only tasked to learn the representation of the density profiles without regard to their explodability. We truncate the density profiles and consider only mass coordinates between $M_{\rm min}$ = 1\,$M_{\odot}$ and $M_{\rm max}$ = 2.3\,$M_{\odot}$.  Interior to $M_{\rm min}$, matter collapses onto the proto-neutron star and lies interior to the stalled shock surface. On the high mass end, it is rare for relevant interfaces in the studied ZAMS distribution to exist beyond $M_{\rm max}$ and still accrete on relevant timescales for neutrino-driven CCSNe.
The density profiles of the \textsc{Kepler} models  with different ZAMS masses (\citealt{swbj16,sukhbold2018}) used as F{\sc{ornax}} supernova progenitors vary in grid resolution between $M_{\rm min}$ and $M_{\rm max}$, ranging typically from 800 to 1200 zones. To standardize the dimension of the input profiles, we interpolate and re-bin the logarithm of the truncated density profiles onto a uniform linear mass grid with $N_{m} = 128$ points. The reduced dimension of 128 is adequate in capturing the sharp jumps in density in the progenitor density profiles (see the solid lines in the top panels of Figure \ref{fig:ae_reconstructions}). To isolate trends, we subtract the means from the profile segments and normalize them independently. Mathematically, for each progenitor, the normalized density profiles take the form:
\begin{align}
 \hat{x}_{m} = (x_{m} - \langle x_{m} \rangle) / (\max(x_{m}) - \min(x_{m}))\,,
\end{align}
where $m$ is the integer index of the uniform mass grid, $x_{m} = \log_{10}(\rho_{m})$ is the logarithmic mass density of the $m$-th mass bin, and $\langle \cdot \rangle$ denotes the mean value over the mass grid. The 128-dimension density profile segments \{$\hat{x}_{m}$\} are used as inputs to the auto-encoders. Mean squared error (MSE) between the input and the reconstructed density profiles \{$\tilde{x}_{m}$\} is used as the loss function for training: $L_{\rm AE} = \sum_{m}(\hat{x}_{m} - \tilde{x}_{m})^2/N_{m}$.

Weights of the convolutional layers are initialized using the \texttt{kaiming\_normal\_} initializer \citep{He15} in \textsc{pytorch}. Biases are initialized as zeros. Optimization is done using the \textsc{Adam} optimizer \citep{KB14}. Training is conducted with a constant batch size of 100 and a learning rate of $10^{-2}$ for 500 epochs.
Since our key goal is to present the utility of physics-agnostic features, we did not perform a systematic hyperparameter study to optimize the auto-encoder.
Due to the limited size of the labeled dataset, we also did not attempt to train the auto-encoder simultaneously with a binary classifier for explosion outcome prediction. These will be instructive follow-up studies when more comprehensive datasets are available.

\subsection{Classifier Training}
\label{sec:classifier}
%Describe the simple workhorse of random forest. Quote the parameters of the random forest: number of estimator, depth, etc.

Using the features obtained in Section \ref{sec:physics_features} and \ref{sec:unsupervised_features}, we train explosion outcome predictors in the form of binary classifiers. We adopt the \textsc{sklearn} implementation of \textsc{RandomForestClassifier} as a common classifier baseline. During training, we adopt a 5-fold, 80/20 split to divide the labeled dataset (with 100 models) into training and testing sets. The training/testing partitions are generated using the \textsc{StratifiedKFold} function of the \textsc{sklearn} package. A constant seed is used for the 5-fold random split, resulting in the same partitions of training/testing data across classifiers trained with different feature sets.
To allow fair comparisons between classifiers trained with different feature sets, we use a relatively simple RF setup with fixed parameters of: \texttt{n\_estimators = 5}, \texttt{criterion = {`gini'}}, \texttt{max\_depth = 3}, \texttt{min\_smaple\_leaf = 2}, and \texttt{max\_features = {`sqrt'}}. The RF classifiers therefore all have a fixed number of five decision trees. With the feature sets we explored in Section \ref{sec:pred_results}, our RF parameters lead to about 7$-$10 nodes per tree partitioning the feature spaces, or about 35-50 total nodes. We use the accuracy, precision, recall, and the F1 score to assess the performance of the classifiers.

\subsection{Semi-supervised Learning with Label Propagation}
\label{sec:lp}

During training, classifiers often require sufficiently large datasets to sample the distributions of various object classes in the feature space. Labeled datasets, in our case multi-dimensional simulations, are costly to produce both in computer and human hours. Alternately, unlabeled datasets, in our application 1D progenitor models, are usually much cheaper to obtain. 
Semi-supervised learning is a hybrid approach devised to use a limited sample of labeled data to assign mock labels to a larger, unlabeled dataset based on some distance metrics in the feature space. The hope is that by propagating the labels to the larger unlabeled dataset and incorporating it in training, the classifier can better learn the data distribution and achieve higher overall prediction accuracy.

Semi-supervised approaches rely on a key presumption that the distributions of different classes are continuous, i.e., data samples close together in the feature space are likely to be of the same class. However, as we have seen in Figure \ref{fig:Ertl_eta_plane} and \ref{fig:SiO_param_plane}, there are complex branches and overlaps associated with degeneracy and/or modeling stochasticity.
Nevertheless, in this paper we attempt the label propagation technique to probe the potential utility of semi-supervised learning approaches in improving explosion outcome prediction.

The procedure of our label propagation study is as follows.  With each feature set selection and in each cross-validation split, we repeat the RF classifier training described in Section \ref{sec:classifier} after (i) randomly removing 50\% or 75\% of the known explosion outcomes in the training split (sized 80) and (ii) re-labeling them based on the distances to their neighbors whose explosion outcomes are retained. In other words, we pretend that our labeled dataset is 50/75\% smaller than it is and allow the label propagation algorithm to re-create a 80-model training set for the RF classifier. Evaluation of prediction performance is still done using the 20-model testing splits.

We employ the \textsc{LabelSpreading} model in \textsc{sklearn} for this task. Fundamentally, \textsc{LabelSpreading} works by building a fully-connected graph connecting all the data samples and propagating labels based on the pairwise distances.
To limit the scope of this exercise, we adopt a constant set of label propagation parameters. In particular, we use the K-nearest neighbor kernel as the distance metric (\texttt{kernel="knn"}) with \texttt{n\_neighbors = 5}. A soft clamping factor of \texttt{alpha = 0.1} is used to allow the algorithm to change at most 10\% of the retained labels from the samples to account for the stochasticity in the explosion simulations.
Pseudo-labels are assigned to the samples with their explosion outcomes removed via the \texttt{transduction\_} operation.

\section{Results}
\label{sec:results}
% baseline classification with physics features, list results in Table 2.
% insights from using the physics-based features.
% then, compare with unsupervised features performance, also list results in Table 2. Key idea: we achieve similar RF performance without human-selected features
% potential directions: look into smaller embedding size, t-SNE analysis, approach for generating new parameters?

\begin{table*}
  \centering
  \begin{tabular}{ccccc}
  \hline \hline
{Features} &  {Accuracy} & Precision & Recall & F1 Score \\
\hline
  $\langle x_{m} \rangle$, $\sigma_{x}$ & 0.68 $\pm$ 0.12 & 0.68 $\pm$ 0.13 & 0.69 $\pm$ 0.13 & 0.67 $\pm$ 0.12 \\
  $\eta_{\rm 1.75}$  & 0.83 $\pm$ 0.08 & 0.84 $\pm$ 0.06 & 0.86 $\pm$ 0.06 & 0.83 $\pm$ 0.07 \\
  $\mu_{4}$, $M_{4} \mu_{4}$  & 0.70 $\pm$ 0.09 & 0.69 $\pm$ 0.08 & 0.69 $\pm$ 0.08 & 0.69 $\pm$ 0.08 \\
  $M_{\rm SiO}$, $\Delta \rho_{\rm SiO}$ & 0.89 $\pm$ 0.10 & 0.89 $\pm$ 0.09 & 0.91 $\pm$ 0.08 & 0.89 $\pm$ 0.10 \\
  Auto-encoder ($d_{z} = 2$) & 0.77 $\pm$ 0.07 & 0.79 $\pm$ 0.07 & 0.75 $\pm$ 0.07 & 0.74 $\pm$ 0.07 \\
  Auto-encoder ($d_{z} = 8$) & 0.84 $\pm$ 0.06 & 0.84 $\pm$ 0.07 & 0.83 $\pm$ 0.06 & 0.83 $\pm$ 0.06 \\
  \hline
  \end{tabular}
  \caption{Table summarizing the performance scores of explosion outcome prediction using different feature sets. Each row corresponds to a different selection of feature parameter(s) used in the training and evaluation of the classifiers. Errors shown are the standard deviations of the respective scores.}
\label{tab:class_perf_global}
\end{table*}

\subsection{Auto-encoder Performance}
\label{sec:ae_results}

Even with the highly limited number of only 106 trainable parameters, the auto-encoders with different embedding dimensions converge efficiently to an MSE loss of $10^{-4} - 10^{-3}$ within less than 50 epochs.
To visualize the representation performance of the auto-encoders, we compare examples of the input and the reconstructed density profiles in the top panels of Figure \ref{fig:ae_reconstructions}. 
With $d_{\rm z} = 2$, the reconstructed profiles miss some of the sharper interface transitions, but trace the overall trends of the profiles quite well. With $d_{\rm z} \ge 8$, the density profiles are all well-captured by the auto-encoders.
In the bottom panels of Figure \ref{fig:ae_reconstructions}, we show the (reconstruction $-$ input) error from all the 100 models in the labeled dataset.
The maximum error is about 0.1 for the auto-encoder with an embedding size of $d_{\rm z} = 2$, whereas for $d_{\rm z} = 8$ the typical deviations are less than about $0.05$.
We emphasize that the representation performance of the auto-encoder architecture can likely be improved or fine-tuned with a more thorough study.
To establish the efficacy of the physics-agnostic features, we choose $d_{\rm z} = 8$ as the fiducial auto-encoder and report the prediction performance in the next section.

\begin{table*}
  \centering
  \begin{tabular}{cccccc}
  \hline \hline
{Features} & Fully Supervised & \multicolumn{2}{c}{50\% Dropped} & \multicolumn{2}{c}{75\% Dropped} \\
 & & No LP & LP & No LP & LP \\
\hline
% $\langle$ $x_{m}$ $\rangle$, $\sigma_{x}$ & - & - & - & - & - \\ 
 $\langle x_{m} \rangle$, $\sigma_{x}$ & 0.68 $\pm$ 0.12 & 0.65 $\pm$ 0.14 & 0.71 $\pm$ 0.14 & 0.61 $\pm$ 0.06 & 0.63 $\pm$ 0.08\\
$\eta_{\rm 1.75}$        & 0.83 $\pm$ 0.08 & 0.81 $\pm$ 0.07 & 0.78 $\pm$ 0.07 & 0.77 $\pm$ 0.10 & 0.74 $\pm$ 0.12 \\
$\mu_{4}$, $M_{4} \mu_{4}$    & 0.70 $\pm$ 0.09 & 0.68 $\pm$ 0.11 & 0.67 $\pm$ 0.08 & 0.62 $\pm$ 0.08 & 0.65 $\pm$ 0.04 \\
$M_{\rm SiO}$, $\Delta \rho_{\rm SiO}$  & 0.89 $\pm$ 0.10 & 0.84 $\pm$ 0.07 & 0.89 $\pm$ 0.09 & 0.82 $\pm$ 0.07 & 0.84 $\pm$ 0.12 \\
Auto-encoder ($d_{z} = 2$) & 0.77 $\pm$ 0.07 & 0.76 $\pm$ 0.04 & 0.72 $\pm$ 0.07 & 0.71 $\pm$ 0.09 & 0.69 $\pm$ 0.06 \\
Auto-encoder ($d_{z} = 8$) & 0.84 $\pm$ 0.06 & 0.74 $\pm$ 0.15 & 0.83 $\pm$ 0.09 & 0.71 $\pm$ 0.10 & 0.73 $\pm$ 0.06 \\
  \hline
  \end{tabular}
  \caption{Table comparing the accuracy scores of different feature sets with 50\% or 75\% of the labels dropped in the RF classifier training set. The Fully Supervised column corresponds to the accuracy using the full labeled training sets (Table \ref{tab:class_perf_global}). The `no LP' columns list the prediction performance trained directly from the reduced-size training set, while the `LP' columns show the performance with label propagation applied to the samples with the labels dropped.}
\label{tab:lp_results}
\end{table*}

\subsection{Explosion Outcome Prediction}
\label{sec:pred_results}

We choose four sets of physics-based features for the explosion outcome prediction task, as listed in the left column of Table \ref{tab:class_perf_global}. The first feature set ($\langle x_{m} \rangle$, $\sigma_{x}$) are the mean and standard deviation of the truncated logarithmic density profiles, representing the most basic summary statistics of the density profiles. The remaining three sets correspond to the physics-based features described in Section \ref{sec:physics_features}.

We train and evaluate a series of RF classifiers using our labeled dataset. Performance scores are listed in Table \ref{tab:class_perf_global} with the errors denoting the standard deviations over the five cross-validation splits. We find that our Si/O interface parameter set provides the best prediction of explosion outcome, with a prediction accuracy of 0.89. By comparison, $\eta_{\rm 1.75}$ yields an accuracy of 0.83 and the Ertl condition 0.70.
Due to the limited size of the labeled set, there are $\approx$10\% fluctuations in the performance scores between different cross-validation splits as well as among repetitions of the RF training (from the randomness in tree-building). However, the general trend of performance with different feature sets is robust.

The embedding feature vector extracted by the fiducial auto-encoder gives a prediction accuracy of 0.84, comparable to both the Si/O interface and the compactness parameter. Even with a reduced embedding dimension of $d_{\rm z} = 2$, the auto-encoder feature vector still outperforms the Ertl parameters.
It highlights that \emph{features obtained from the density profiles via an unsupervised, physics-agnostic manner can offer classification performance competitive with physics-based features}.

Identifying a prominent Si/O interface can be difficult, particularly with low-mass models, perhaps explaining some of the misclassifications. We find that stars between 12$-
$15 M$_{\odot}$ lack prominent Si/O interfaces and tend to be more difficult to explode. According to \cite{sukhbold2018}, stars in this range may have multiple smaller, fragmented interfaces as well. We emphasize that our conclusion depends sensitively on both the progenitor profile and the neutrino microphysics included (see for instance, \citealt{burrows_2019}). Regardless, multiple studies have indeed found that models generally within this mass range are less likely to explode \citep{burrows_2019,vartanyan2018a, oconnor_couch2018a, summa2016, burrows_2019, 2021Natur.589...29B,2022arXiv220702231W}.

The accuracy of the compactness parameter is surprising at first sight, given that studies have found no simple correlation between compactness and explosion outcome \citep{burrows2020,2021Natur.589...29B}. Indeed, we see no monotonic dependence of explosion outcome on compactness in Figure \ref{fig:Ertl_eta_plane}. Rather, the RF classifier here identifies the non-linear mapping between compactness and explosion outcome from the training samples.
Such a non-linear dependence of explosion outcome is suggestive of additional underlying physics that is not captured by the compactness parameter, perhaps in the nuances of the density profile.

Regardless of the metric used, we find predictive accuracy above 70\%, indicating that all the metrics considered contain some physical information about the explosion outcome. 
The Ertl parameter just underperforms compared to simply using only the density and its standard deviation. While both the Ertl parameter and compactness contain information about the density of the progenitor, the former may obfuscate it through analytical complication, while the latter, which performs better, is oversimplified. Categorizing the density profile through prominent interfaces seems to be the best approach thus far to predicting explosion outcome.

\subsection{Utility of Label Propagation}
\label{sec:label_propagation}

Table \ref{tab:lp_results} summarizes the results of the label propagation study. Unsurprisingly, classifiers trained with fewer labeled samples tend to have poorer prediction performance across feature sets. Even with 75\% of the labeled training samples dropped, i.e., only with 20 training samples, the classifiers can still preserve reasonable prediction accuracy scores of about 0.6 $-$ 0.8 (the `no LP' columns). This suggests that the feature spaces we investigated can be sampled reasonably well with about 20 $-$ 40 models, and that most of the mis-classifications reside in the overlaps of branches that may be resolved by additional training samples.

Label propagation offers only marginal improvements of a few percentage points across different feature sets. In some cases, e.g., with $\eta_{1.75}$ and auto-encoder features of embedding dimension $d_{\rm z} = 2$, label propagation can even diminish prediction performance. Such reduction in accuracy can be understood again by the complex discontinuities in the feature spaces and the stochasticity in modeling. With a small number of training samples, the classifiers can sometimes be misled by a single training sample to mis-classify large parts of the feature space. 
With a much larger dataset, the overlapping outcome branches will be better distinguished. We expect label propagation to be more effective in improving prediction accuracy with feature spaces that are smoother and less susceptible to model stochasticity. The unsupervised approach of feature extraction holds promise in uncovering such feature spaces.

\section{Conclusions}\label{sec:conc}

We explored the utility of a machine learning framework in predicting the explosion outcomes of massive stars based on their 1D progenitor models. We trained and evaluated a basic random forest classifier as an explosion predictor using both physics-based and physics-agnostic features. In particular, we investigated the commonly used compactness parameter and Ertl conditions, a new feature set that quantifies the location and the extent of the density drop at the silicon/oxygen interface, and auto-encoder features generated from the progenitor density profiles in an unsupervised manner.  Applied to a set of 100 2D radiation hydrodynamical F{\sc{ornax}} simulations, we found that the new silicon/oxygen interface feature set has the best predictive power, with an accuracy of $\approx$90\%, outperforming the compactness and the Ertl condition. More importantly, using the physics-agnostic auto-encoder features, we obtained a predictive accuracy of $\approx$84\%, second only to the silicon/oxygen interface features.

The competitive predictive performance of the auto-encoder features revealed that the density profiles alone contain meaningful information about the explodability of the stellar progenitors. It suggests that exploration of the clusters and branches in the reduced-dimension embedding space holds promise in uncovering the underlying progenitor properties that foreshadow explosions. With more multi-dimensional explosion simulations in the near future, we expect the unsupervised approach to representing progenitor models to be profitable in the task of identifying more robust explosion physics.

\section{Caveats and Future Prospects}
\label{sec:future}

% uncertainties in modeling physics
% improvements on ML approach b/c parameter space is large
% limited dataset size
% future directions beyond predicting outcome

The conclusions cited here assume neutrino-heating as the dominant explosion mechanism, as is well-understood to be the case for the majority of garden-variety CCSNe. Explosion outcome depends on a confluence of factors (details of the code, simulation dimensions, stochasticity) and physical uncertainties (microphysics, progenitor structure, convection, nuclear burning rates). 
Our model suite can be extended to include additional physics, including magnetorotational effects, for instance, to capture more of the physical parameter space yet unexplored in CCSNe simulations.

Importantly, the stellar progenitors used were all spherically-symmetric, one-dimensional models. Only recently (\citealt{2016ApJ...833..124M,zha2019, takahashi2019,fields2020,fields2021,fields2022}) have multi-dimensional progenitor models become available for simulation (\citealt{muller:18, muller_lowmass, 2022MNRAS.510.4689V,2022MNRAS.513.1317Z}). CCSNe simulations are sensitive to the ambient perturbations in the progenitor model (see, e.g. \citealt{burrows_2019}). The structure of the prominent compositional interfaces, and hence explosion outcome, morphology, and nucleosynthetic yields, will differ between 3D and 1D progenitor models.

The relevant physical parameter space for explosion outcome is both very large and poorly constrained. Even the presence of a strong interface is difficult to resolve, and sometimes absent, in many progenitors. Additional constraints, perhaps involving the Helium core mass or some other characterization of the density profile (\citealt{sukhbold2018,2022arXiv220702231W}) is needed to break the degeneracy in predicting explosion outcome.
We focused exclusively on the density profile when computing both the physics-based and physics-agnostic auto-encoder features,
but we could expand the feature sets to include also the temperature, electron-fraction, and/or entropy profiles, etc. from the progenitor stellar models. For example, multiple profiles can be readily incorporated as different input channels in the auto-encoder architecture.
Our main goal is to demonstrate the usefulness of the unsupervised feature extraction approach. We therefore did not conduct a thorough hyperparameter study for the auto-encoder architecture.
Exploring the utility of transformers \citep{Vaswani17}, another neural network architecture that is effective in representing sequential data, will also be a promising future direction. 

Furthermore, our dataset was limited in size. We selected from approximately 1500 progenitor models and trained on 100 axisymmetric 2D simulations. Although explosion outcome does not seem to differ greatly between 2D and 3D simulations (\citealt{vartanyan2018b, 2021Natur.589...29B}), a significantly larger catalog of simulations, even in axisymmetry, would better populate the distribution of density profiles by progenitor, perhaps better resolving the clustering and outcome branches (\citealt{2016MNRAS.460..742M,sukhbold2018}) seen in the different phase spaces explored here. At the very least, we would need tenfold more simulations (thousands), even in 2D, to have a more balanced and comprehensive dataset. Yet even with the limited dataset and our simple approach in both identifying a physical criterion of interest and apposite ML techniques, we were able to obtain promising results.

Machine learning applications are not limited to a simple binary determination of explosion outcome. Regression models can enable prediction of explosion diagnostics, such as the energy and ejecta composition, for a given physical setup and progenitor model. Inverse modeling can facilitate the reconstruction of progenitor properties from observables. Data transformation and image segmentation, which have already seen some use categorizing observations, can be  used to characterize morphological features of supernova remnants, such as nickel bullets and voids/clustering in the ejecta, which, jointly with inverse modeling can characterize the structure of the stellar progenitor and its evolutionary history. Machine learning techniques provide an invaluable perspective from which to map the initial stellar mass function to the distribution of residues, i.e., black holes and neutron stars (partitioned between failed and successful supernovae), and are exquisitely suitable for the upcoming era of all-sky surveys. The effort here presents a first step in this direction.

\software{
          \textsc{Kepler} \citep{swbj16},
          F{\sc{ornax}} \citep{skinner2019},
          Jupyter \citep{Kluyver16},
          Matplotlib \citep{Hunter07},
          NumPy \citep{Oliphant06},
          \textsc{sklearn} \citep{scikit-learn}.
          }

\section*{Acknowledgements}
We are grateful to Daniel Kasen, Tianshu Wang, and Matthew Coleman for valuable insights and discussion. 
This research was funded by the Gordon and Betty Moore Foundation through Grant GBMF5076, and by NASA awards ATP-80NSSC18K0560 and ATP-80NSSC22K0725.
This research was supported in part by the National Science Foundation under Grant No. NSF PHY-1748958.
We acknowledge support from the U.S. Department of Energy Office of Science and the Office
of Advanced Scientific Computing Research via the Scientific Discovery
through Advanced Computing (SciDAC4) program and Grant DE-SC0018297
(subaward 00009650) and support from the U.S. NSF under Grants AST-1714267
and PHY-1804048 (the latter via the Max-Planck/Princeton Center (MPPC) for Plasma Physics).
A generous award of computer time was provided
by the INCITE program. That research used resources of the
Argonne Leadership Computing Facility, which is a DOE Office of Science
User Facility supported under Contract DE-AC02-06CH11357. We are also grateful for our computational resources through the Texas Advanced Computing Center (TACC) at The University of Texas at Austin via Frontera Large-Scale Community Partnerships under grant SC0018297 as well as the Leadership Resource Allocation under grant number 1804048. In addition, this overall research
project was part of the Blue Waters sustained-petascale computing project,
which is supported by the National Science Foundation (awards OCI-0725070
and ACI-1238993) and the state of Illinois. Blue Waters was a joint effort
of the University of Illinois at Urbana-Champaign and its National Center
for Supercomputing Applications. This general project was also part of
the ``Three-Dimensional Simulations of Core-Collapse Supernovae" PRAC
allocation support by the National Science Foundation (under award \#OAC-1809073).
Moreover, we acknowledge access under the local award \#TG-AST170045
to the resource Stampede2 in the Extreme Science and Engineering Discovery
Environment (XSEDE), which is supported by National Science Foundation grant
number ACI-1548562. Finally, the authors employed computational resources provided by the TIGRESS high
performance computer center at Princeton University, which is jointly
supported by the Princeton Institute for Computational Science and
Engineering (PICSciE) and the Princeton University Office of Information
Technology, and acknowledge our continuing allocation at the National
Energy Research Scientific Computing Center (NERSC), which is
supported by the Office of Science of the US Department of Energy
(DOE) under contract DE-AC03-76SF00098.
Use was made of computational facilities purchased with funds from the National Science Foundation (CNS-1725797) and administered by the Center for Scientific Computing (CSC). The CSC is supported by the California NanoSystems Institute and the Materials Research Science and Engineering Center (MRSEC; NSF DMR 1720256) at UC Santa Barbara.

%%%%%%%%%%%%%%%%%%%%%%%%%%%%%%%%%%%%%%%%%%%%%%%%%%
\section*{Data Availability}

The data underlying this article will be shared on reasonable request to the corresponding author.

\bibliographystyle{aasjournal}
\bibliography{References}

\ifMNRAS
\bsp	% typesetting comment
\label{lastpage}
\else
\fi

\end{document}